\documentclass[aps,twocolumn,showpacs]{revtex4}
\usepackage{amssymb}
\usepackage{graphicx}
\usepackage{bm}
\usepackage{amsmath}

\newcommand{\beq}{\begin{equation}}
\newcommand{\eeq}{\end{equation}}
\newcommand{\beqn}{\begin{eqnarray}}
\newcommand{\eeqn}{\end{eqnarray}}

\begin{document}

\title{Chiral gauge field and axial anomaly in a Weyl semi-metal}
\author{Chao-Xing Liu$^1$, Peng Ye$^2$ and Xiao-Liang Qi$^3$}
\affiliation{$^1$ Department of Physics, The Pennsylvania State University, University Park,
Pennsylvania 16802-6300\\
$^2$ Institute for Advanced Study, Tsinghua
University, Beijing, 100084, People's Republic of China\\
$^3$ Department of
Physics, McCullough Building, Stanford University, Stanford, CA
94305-4045}
\date{\today}

\begin{abstract}
Weyl fermions are two-component chiral fermions in $(3+1)$-dimensions. When coupled to a gauge field, the Weyl fermion is known to have an axial anomaly, which means the current conservation of the left-handed and right-handed Weyl fermions cannot be preserved separately. Recently, Weyl fermions have been proposed in condensed matter systems named as ``Weyl semi-metals". In this paper we propose a Weyl semi-metal phase in magnetically doped topological insulators, and study the axial anomaly in this system. We propose that the magnetic fluctuation in this system plays the role of a ``chiral gauge field" which minimally couples to the Weyl fermions with opposite charges for two chiralities. We study the anomaly equation of this sytem and discuss its physical consequences, including one-dimensional chiral modes in a ferromagnetic vortex line, and a novel plasmon-magnon coupling.

\end{abstract}

\pacs{  } \maketitle

{\it Introduction - }
In the quantum field theory, a (3+1)-dimensional massless Dirac fermion is decomposed to two independent two-component fermions known as the Weyl fermions. Weyl fermion has a definite chirality, lefthanded or righthanded, determined by the sign of its spin polarization along the momentum direction.\cite{peskin1995}
Classically, the lefthanded and righthanded Weyl fermions are decoupled and can be coupled independently to two gauge fields, leading to a separate charge conservation. The gauge field that couples differently to Weyl fermions with two chiralities is called a chiral gauge field. For example the SU(2) gauge field in the Standard Model is a chiral gauge field. It is well-known that the chiral charge conservation is violated in a quantum theory of Weyl fermions in a background gauge field, which is known as the axial anomaly\cite{adler1969,bell1969,nielsen1983}.

Recently, Weyl fermions are also introduced into condensed matter physics. The Weyl fermions are shown to be the topologically robust boundary states of $(4+1)$-d time-reversal (TR) invariant topological insulators (TI)\cite{qi2008a}, and the axial anomaly corresponds to a topological response of the $(4+1)$-d TI. This approach is related to the domain wall fermion approach\cite{kaplan1992} and Callan-Harvey effect\cite{callan1985} in high energy physics. By dimensional reduction, the $(4+1)$-d topological insulator is reduced to the $(3+1)$-d TI\cite{qi2011,hasan2010,moore2010} and the Weyl fermion is reduced to $(2+1)$-d surface states of the TI. Weyl fermions also appear directly in $(3+1)$-d gapless electron systems, which are named as ``Weyl semi-metals"\cite{volovik2003,wan2011,xu2011,burkov2011a,burkov2011b,cho2011,fang2011,jiang2011,hosur2012,yang2011,nielsen1983,aji2011}. Since a system with both TR and parity (P) symmetries have all energy bands doubly degenerate, the Weyl semi-metal state can only be realized in a system breaking TR and/or P symmetry.

A natural question is whether the chiral gauge field can be realized in the Weyl semi-metals, and if yes, what is the physical consequence.
In this letter, we address these questions in TR breaking Weyl semi-metals. We show that generically a ferromagnetic moment couples to the Weyl fermions as a chiral gauge field. As an explicit example system, we study a model of magnetically doped topological insulator, which can be driven into the Weyl semi-metal phase with strong enough magnetic moments. The presence of the chiral gauge field leads to an anomaly equation satisfied by the charge current, which leads to new topological phenomena such as chiral one-dimensional states in a magnetic vortex, and a topological coupling between spin fluctuation and plasmons. 

{\it Chiral gauge field and anomaly equation - }
We start with a general discussion of Weyl fermions in condensed matter physics. 
In a weakly interacting crystalline material, Weyl fermion states generically appear when two energy bands cross at a generic point $\vec{K}_0$ in the Brillouin zone. The low energy physics around $\vec{K}_0$ is described by a two-component Hamiltonian
$H_W=\hbar \sum_{i,j=x,y,z}v_{ij}k_i\sigma_j$, with $k_i$ the momentum away from $\vec{K}_0$, and $\sigma_j$ the Pauli matrices. The matrix $v_{ij}$ describes the generic linear coupling between momentum and spin degree of freedom described by $\sigma_j$. By rotating the basis one can always diagonalize $v_{ij}$, and the three diagonal components are anisotropic velocities. Without losing generality, we restrict our discussion on isotropic Weyl fermions with the simple Hamiltonian $H=\hbar v_f{\vec{ \sigma}\cdot \vec{k}}$. Our results on anomaly and chiral gauge field is insensitive to the anisotropy in the velocity. The sign of the Fermi velocity $v_f$ determines the chirality of Weyl fermion. 
A single Weyl fermion is topologically stable as long as translation symmetry is preserved\cite{wan2011}, thus for a Weyl semi-metal with translation symmetry,
any local perturbation can only move the nodal point $\vec{K}_0$ in the momentum space.

According to the Nielsen-Ninomiya theorem \cite{nielsen1981a,nielsen1981b}, in a lattice model the number of Weyl fermions with opposite chiralities must be equal. Consequently, the minimum number of Weyl fermions in a Brillioun zone is 2. Moreover, because TR symmetry preserves the chirality of Weyl fermion, in TR invariant system the minimum number of Weyl fermions is 4\cite{burkov2011b}.
In the following, we focus on the ``minimal Weyl semi-metal" which break TR but preserves P, with two Weyl fermions of opposite chiralities at wavevectors $\vec{K}_0$ and $-\vec{K}_0$, related to each other by spatial inversion. $\vec{K}_0$ is a generic point in BZ away from TR invariant momenta. 

We consider an arbitrary perturbation to the system of two Weyl fermions. As long as the perturbation is so smooth that the momentum transfer is much smaller than $2|\vec{K}_0|$, the two Weyl fermions remains decoupled. The effective Hamiltonian of the lefthanded Weyl fermion under perturbation is $H_L=\hbar v_f\vec{\sigma}\cdot\vec{k}+\delta H_L$, with $\delta H_L$ a generic $2\times 2$ Hermitian matrix. To the leading order one can ignore the $k$ dependence and consider $\delta H_L$ as a constant term. Then $\delta H_L$ can always be expanded to the form $\delta H_L=\hbar v_f\vec{\sigma}\cdot \vec{a}_L+a_{0L}$ with the last term proportional to identity. Adding this to the Weyl fermion Hamiltonian we find $H_L=\hbar v_f(\vec{k}+\vec{a}_L)\cdot\vec{\sigma}+a_{0L}$ with $a_{\mu L}=(a_{0L},\vec{a}_L)$ behaving as a gauge field. Similarly one can define the gauge field $a_{\mu R}$ minimally coupled to the righthanded Weyl fermions with the Hamiltonian $H_R=-\hbar v_f(\vec{k}+\vec{a}_R)\cdot\vec{\sigma}+a_{0R}$.  
The two Weyl fermions can be described together by a $4\times 4$ Hamiltonian:
\begin{eqnarray}
	&&H
        =\hbar v_f\left( (\vec{k}+\vec{A})\cdot\vec{\sigma}\tau_z+\vec{a}\cdot\vec{\sigma}\right)+a_0\tau_z +A_0
	\label{eq:CGF_Ham1}
\end{eqnarray}
with $A_\mu=(a_{\mu L}+a_{\mu R})/2$ behaving like the electromagnetic gauge field, and $a_\mu=(a_{\mu L}-a_{\mu R})/2$ the chiral gauge field. 
$A_\mu$ and $a_\mu$ have different properties under P and TR. For example if the perturbation we consider is a fluctuation of a ferromagnetic moment, only $\vec{a}$ will be induced which is TR odd and P even.

As known from the quantum field theory, when a Weyl fermion is coupled to a gauge field, the charge conservation is broken
at the quantum field level, leading to the axial anomaly\cite{peskin1995,zee2010}, which can be described by the anomaly equation
$\partial_{\mu}j^{\mu L(R)}=(-)\frac{1}{32\pi^2}\epsilon^{\lambda\rho\mu\nu}f^{L(R)}_{\lambda\rho}f^{L(R)}_{\mu\nu}$ where
$f^{L(R)}_{\mu\nu}=\partial_{\mu}a_{\nu L(R)}-\partial_{\nu}a_{\mu L(R)}$. Since the gauge field $a_{\mu L(R)}$ of lefthanded (righthanded) Weyl fermion
is related to the gauge field $A_{\mu}$ and $a_{\mu}$, the anomaly equation can also be rewritten as
$\partial_\mu j^{\mu L(R)}=(-)\frac{1}{32\pi^2}\epsilon^{\lambda\rho\mu\nu}(F_{\lambda\rho}+(-)f_{\lambda\rho})(F_{\mu\nu}+(-)f_{\mu\nu})$,
where $F_{\mu\nu}=\partial_{\mu}A_{\nu}-\partial_{\nu}A_{\mu}$ is electromagnetic field strength and
$f_{\mu\nu}=\partial_{\mu}a_{\nu}-\partial_{\nu}a_{\mu}$ is chiral gauge field strength.
Let's define the charge current as $j^{\mu}=j^{\mu L}+j^{\mu R}$ and the axial current as $j^{\mu 5}=j^{\mu R}-j^{\mu L}$.
When $\vec{A}$ and $\vec{a}$ coexist, we find both the axial current and charge current are nonconserved with the anomaly equations
\begin{eqnarray}
	&&\partial_\mu j^{\mu 5}=-\frac{1}{16\pi^2}\epsilon^{\lambda\rho\mu\nu}(F_{\alpha\beta}F_{\mu\nu}+f_{\lambda\rho}f_{\mu\nu}), \label{eq:CGF_Anomalyaxial} \\
	&&\partial_\mu j^{\mu}=\frac{1}{8\pi^2}\epsilon^{\lambda\rho\mu\nu}f_{\lambda\rho}F_{\mu\nu}.
	\label{eq:CGF_Anomaly}
\end{eqnarray}
The equation (\ref{eq:CGF_Anomalyaxial}) is the axial current anomaly\cite{nielsen1983,aji2011} but with the additional term induced by chiral gauge field,
while the equation (\ref{eq:CGF_Anomaly}) indicates the conservation of charge current is also broken
due to the combination effect of chiral gauge field and electromagnetic field, which is the main focus of this paper.
At the first glance, the breaking of the charge conservation seems something unphysical. Here we emphasize that the Weyl fermion description is only
a low energy effective theory and the high energy part is not taken into account.
Let's define $j_b^{\mu}=-\frac{1}{2\pi^2}\epsilon^{\mu\nu\lambda\rho}a_{\nu}\partial_{\lambda}A_{\rho}$,
and the right-hand side of equation (\ref{eq:CGF_Anomaly}) can be written as a total derivative of $j^{\mu}_b$, and the charge conservation law
$\partial_{\mu}\left( j^{\mu}+j_b^{\mu} \right)=0$ is recovered if $j^{\mu}_b$ is regarded as a current from the high energy part which is neglected in our description.
Actually we notice that the spatial component of $j^{\mu}_b$ is given by $\vec{j}_b=-\frac{1}{2\pi^2}\vec{a}\times \vec{E}$ with the electric field $\vec{E}$,
exactly corresponding to the anomalous Hall response of Weyl fermion, as first derived in Ref \cite{yang2011}.
To make our discussion concrete, we first propose a realization of Weyl fermions and chiral gauge field in magnetically doped topological insulators, before discussing the physical consequence of this anomaly equation. 

\begin{figure}[tbp]
\begin{center}
\includegraphics[width=3.5in]{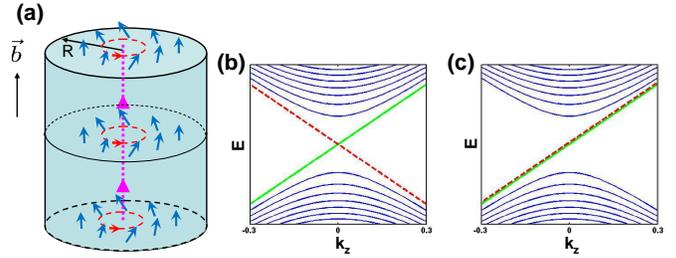}
\end{center}
\caption{ (a) ``Chiral magnetic field'' can be generated by the magnetic vortex configuration in a topological insulator cylinder.
Here the vector $\vec{b}$ indicates the direction of the ``chiral magentic field''.
The Landau level spectrum of a massless Dirac fermion is plotted (b) for a uniform magnetic field $\vec{B}$
and (c) for a uniform ``chiral magentic field'' $\vec{b}$. } \label{fig:vortex}
\end{figure}

\begin{figure}[tbp]
\begin{center}
\includegraphics[width=3.5in]{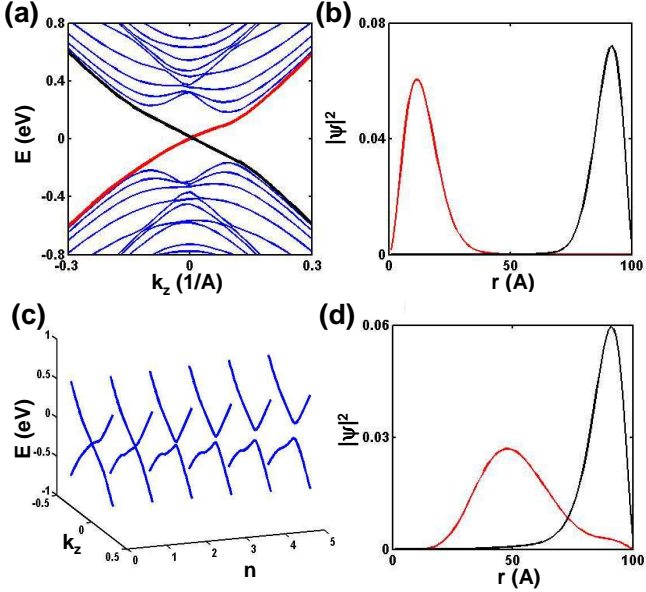}
\end{center}
\caption{ (a) The energy dispersion as a function of $k_z$ along a ferromagnetic vortex line in a topological insulators
with the total angular momentum $J_z=\frac{1}{2}$.  (b) The corresponding radial wave function for the two chiral modes at $k_z=0$.
Here red line is for the wave function near the vortex core (r=0) and the black line for the one at r=R.
(c) The energy dispersion of the two zero modes for different $J_z=n+\frac{1}{2}$. The gap is observed for large $|n|$, due
to the finite size effect, which gives a cut-off of the total number of chiral modes. In (d),
the wave function of the inner chiral mode moves outwards, hybridizing with the outer chiral mode (Here we take $n=3$ and $k_z=0.06$ 1/\AA).
The parameters of the four band model are taken to be $M_0=0$, $M_1=0.342eV\cdot$\AA$^2$, $M_2=18.25eV\cdot$\AA$^2$, $B_0=1.33eV\cdot$\AA,
$A_0=2.82eV\cdot$\AA, $U_0=0.1eV$ and $W_0=0.06eV$. 
} \label{fig:chiral}
\end{figure}

{\it Material realization -}
It is first suggested that Weyl fermions can be realized in pyrochlore iridates\cite{wan2011}, and later another material HgCr$_2$Se$_4$ is also proposed\cite{xu2011}.
However both the materials include multiple Weyl fermions with the number larger than 2, making the system complicated,
therefore it is desired to have a system with the minimal number of Weyl fermions,
which actually can be achieved by magentically doped topological insulators\cite{burkov2011a,cho2011}.
By substituting the atoms, it is possible to tune the band gap of topological insulators, and even induce the phase transition between trivial
and non-trivial phases, which has been realized in TlBi(S$_{1-\delta}$Se$_{\delta}$)$_2$ recently\cite{yan2010,lin2010,xu2011a}. 
Near the transition point, the
bulk gap is minimized and can be overcomed by the exchange coupling from magnetic doping.
The ferromagnetism in the Cr or Fe doped Bi$_2$Te$_3$ and Sb$_2$Te$_3$ has been
observed in experiment\cite{chen2010,chang2011,wray2011}, therefore the magnetically doped Bi$_2$Se$_3$ and TlBiSe$_2$ family of materials are
the suitable platform for the realization of minimal number of Weyl fermions. Here we adopt the four band model\cite{zhang2009,liu2010}
with general mass terms, to describe these materials,
\begin{eqnarray}
	&&H=H_0+H_1\label{eq:MR_Ham1}\\
	&&H_0=\epsilon(\vec{k})+\mathcal{M}(\vec{k})\Gamma_5+B_0k_z\Gamma_4+A_0(k_y\Gamma_1-k_x\Gamma_2)\nonumber\\
	&&H_1=\sum_{ij}m_{ij}\Gamma_{ij}\nonumber
\end{eqnarray}
where $\epsilon_{\bold{k}}=C_0+C_1k_z^2+C_2k^2_\parallel$, $\mathcal{M}(\bold{k})=M_0+M_1k_z^2+M_2k^2_\parallel$.
The $\Gamma$ matrices are defined as $\Gamma_{1,2,3}=\sigma_{x,y,z}\tau_x$,
$\Gamma_4=\tau_y$, $\Gamma_5=\tau_z$, and $\Gamma_{ab}=[\Gamma_a,\Gamma_b]/2i$ ($a,b=1,\dots,5$).
Ferromagnetism breaks T but preserves P, therefore by inspecting the symmetry property of $\Gamma$ matrices (eg. the table III
in the reference \cite{liu2010}), we immediately find only two sets of $\Gamma$ matrices are allowed in $H_1$: $\Gamma_{ij}=\varepsilon_{ijk}\sigma_k$
and $\Gamma_{i4}=\sigma_i\tau_z$ ($i,j,k=x,y,z$). Generally $\Gamma_{12}$ and $\Gamma_{34}$ can be induced
by z-direction magnetization, while $(\Gamma_{14},\Gamma_{24})$ and $\left( \Gamma_{23},\Gamma_{31}\right)$ originate from in-plane magnetization.
It is shown that $\Gamma_{14}$, $\Gamma_{24}$ and $\Gamma_{12}$ induce two Weyl fermions while $\Gamma_{23}$, $\Gamma_{31}$ and $\Gamma_{34}$
yield a nodal ring\cite{burkov2011b}. Since now we are interested in the Weyl fermion regime, we focus on the simple case with $H_1=U_{0}\Gamma_{12}$,
yielding the energy dispersion $E_{st}=\epsilon(\vec{k})+s\sqrt{A_0^2(k_x^2+k_y^2)+(\sqrt{\mathcal{M}^2+B_0^2k_z^2}+t|U_0|)^2}$,
where $s,t=\pm 1$. The two bands in the middle with $t=-1$ touches 
when the conditions $\mathcal{M}^2+B^2_0k^2_z=U_0^2$ and $k_x=k_y=0$ are satisfied. If we 
neglect the quadratic term in $\mathcal{M}$ for simplicity, the bulk gap is closed, realizing Weyl fermions, at the momentum $k_z=\pm K_0$ with $K_0=\frac{1}{B_0}\sqrt{U_0^2-M_0^2}$ if  $|U_0|>|M_0|$.

\begin{figure}[tbp]
\begin{center}
\includegraphics[width=3.5in]{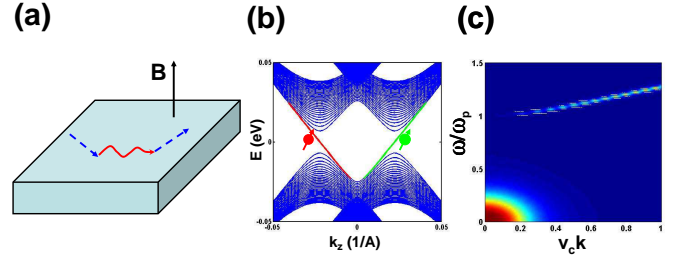}
\end{center}
\caption{ (a) The experiment setup and the diagram for the corresponding physical process.
Here the red line denotes the electromagnetic field, and the blue dashed line is for $a_z$ field.
(b) The Landau levels for the four band model (\ref{eq:MR_Ham1}) in the uniform magnetic field.
(c) The imaginary part of the correlation function $\langle a_z(q)a_z(-q)\rangle$. Here we take $\omega_0/\omega_p=0.1$, $v_s=0$.  } \label{fig:setup}
\end{figure}

Next we consider the perturbation around the gapless points $\vec{k}=(0,0,\pm K_0)$ with the perturbed Hamiltonian
$H'=B_0 \delta k_z\tau_y+A_0\left(  \delta k_y\sigma_x\tau_x- \delta k_x\sigma_y\tau_x \right)
+ \sum_{i=x,y,z}\left(  \mu_i\sigma_i+  \nu_i\sigma_i\tau_z \right)$,
where $\vec{\mu}$ and $\vec{\nu}$ denote the magnetic fluctuation and $\delta \vec{k}$ is the momentum expanded around $(0,0,\pm K_0)$.
We project the Hamiltonian $H'$ into the subspace expanded by $|-,-\rangle$ and $|+,-\rangle$ by perturbation theory,
obtaining the effective Hamiltonian $H_{eff}=\hbar v_{fz}(\delta k_z\sigma_z\tau_z+a_z\sigma_z)+
	\hbar v_{f\parallel}\sum_{i=x,y}\left( \delta k_i\sigma_i\tau_z+a_i\sigma_i \right)$,
where $\sigma$ denotes spin, $\tau$ represents two Dirac cones at $(0,0,\pm K_0)$,
the Fermi velocity $\hbar v_{fz}=-\frac{B_0^2K_0}{U_0}$, $\hbar v_{f\parallel}=-A_0$ and the chiral gauge potential $\vec{a}$ is given by
\begin{eqnarray}
	&&\hbar v_{fz} a_z=\mu_z-\frac{1}{U_0}\left( \nu_z^2-\mu_x^2-\mu_y^2-\frac{M_0^2}{U_0^2}\sum_i\nu_i^2 \right)\label{eq:MR_chiral1}\\
	&&\hbar v_{f\parallel} a_x=\frac{B_0K_0}{U_0}\nu_x-\frac{B_0K_0}{U_0^2}\nu_z\mu_x\label{eq:MR_chiral2}\\
	&&\hbar v_{f\parallel} a_y=\frac{B_0K_0}{U_0}\nu_y-\frac{B_0K_0}{U_0^2}\nu_z\mu_y.
	\label{eq:MR_chiral3}
\end{eqnarray}

{\it Physical consequence -}
Let's introduce ``chiral magnetic field'' $\vec{b}=\nabla\times\vec{a}$ and ``chiral electric field'' $\vec{e}=\frac{\partial\vec{a}}{\partial t}$, then equation (\ref{eq:CGF_Anomaly}) can be rewritten as
\begin{eqnarray}
	\frac{\partial\rho}{\partial t}+\vec{\nabla}\cdot\vec{j}=\frac{1}{2\pi^2}\left( \vec{b}\cdot\vec{E}+\vec{e}\cdot\vec{B} \right),
	\label{eq:PC_Anomaly}
\end{eqnarray}
with $\rho$ and $\vec{j}$ the charge density and current, respectively. In the following we study the physical consequences of the two terms on the righthand side of this equation.

The first term $\vec{b}\cdot\vec{E}$ describes the effect of a chiral magnetic field parallel to the electric field.
The ``chiral'' magnetic field can also induce Landau levels, similar to the Landau levels by magnetic field.
For a uniform field $\vec{b}=b_0\hat{e}_z$, the Landau level spectrum is given by
$E_{\pm,\alpha}(n)=\pm\hbar v_f\sqrt{k_z^2+2b_0n}$ with $n=1,2,\dots$ and $\alpha=\pm$ denote two Dirac cones.
In addition, there are two zeroth Landau levels, both with the dispersion $E_{\alpha}(0)=-\hbar v_fk_z$ ($\alpha=\pm$), as shown in Fig.\ref{fig:vortex} (c). It should be noticed that the two zeroth Landau levels are one-dimensional modes with the same chirality, in contrast to the case of 
an ordinary magnetic field $\vec{B}=B_0\hat{e}_z$ shown in Fig. \ref{fig:vortex} (b). In other words, the low energy dynamics of the system is described by two chiral fermions in each area with one flux quanta of $\vec{b}$. In this case, the anomaly equation (\ref{eq:PC_Anomaly}) reduces to the chiral anomaly of 1D chiral fermions\cite{johnson1963,jackiw1985}.

A key difference of the chiral gauge field from the electromagnetic gauge field is that the gauge vector potential $\vec{a}$ is physical and thus has to be single valued. A uniform $\vec{b}$ field corresponds to a $\vec{a}$ linearly increasing towards the boundary of the system, which is unphysical. Also the Landau level spectrum above does not consider the coupling between the two Weyl fermions. To obtain a more complete understanding to this problem, we consider the four band model (\ref{eq:MR_Ham1}) with the magnetic vortex configuration
$m_{14}=-W_0\sin\theta$, $m_{24}=W_0\cos\theta$ ($\theta$ is the angular coordinate), $m_{12}=U_0$ and all the other $m_{ij}=0$. Assuming a cylinder shape geometry as shown in Fig. (\ref{fig:vortex}) (a), the system has translation symmetry in $z$ direction and rotation symmetry according to $z$ axis. Therefore the momentum $k_z$ and total angular momentum $J_z$ are good quantum numbers, and the Schordinger equation can be solved numerically for each sector of $k_z,J_z$ by introducing a discretization, as described in the appendix in detail. Fig \ref{fig:chiral} (a) and (b) show the band dispersion along z direction for the angular momentum $J_z=1/2$, and the corresponding wave function. There are two gapless modes with opposite chirality, different from the uniform field Landau levels in Fig. \ref{fig:vortex} (c). The wavefunction shows that the two chiral modes are spatially separated, with one wave function around $r=0$ and the other one located at the boundary $r=R$, as shown in Fig \ref{fig:chiral} (a) and (b). From this result we see that the chiral modes in the zeroth Landau level are compensated by modes with opposite chirality on the boundary, which is expected since in such a finite system the number of left and right moving 1D states must be equal.
With increasing the angular momentum $J_z$, the wavefunction of the inner chiral mode moves towards larger radius, as shown in Fig. \ref{fig:chiral} (d). For a finite system, 
the wave functions of the two chiral states near $r=0$ and $r=R$ will overlap with each other for large angular quantum $J_z$ (Fig \ref{fig:chiral} (d)), leading to a gap opening,
as shown in Fig \ref{fig:chiral} (c). 
Consequently we obtain a finite number of chiral modes in the zeroth Landau level. As expected, the number of chiral modes is determined by the total flux of $\vec{a}$ in the system, just like the case of uniform $\vec{b}$ field. In such a configuration, the consequence of the anomaly equation (\ref{eq:PC_Anomaly}) is actually a quantum Hall effect\cite{yang2011}. In an electric field $\vec{E}=E\hat{z}$ parallel to $\vec{b}$, the anomaly equation describes a charge generation around the center of the system, while the charge on the boundary is annihilated. This is a consequence of a Hall current flowing along the radial direction towards the center, which can be measured in transport experiments.

The second term on the right hand side of the anomaly equation (\ref{eq:PC_Anomaly}) describes the combination effect of magnetic field and ``chiral'' electric field. To understand this term, consider a uniform magnetic field $\vec{B}=B_0\hat{z}$ and a uniform vector potential $\vec{a}=a_z(t)\hat{z}$ changing adiabatically in time. The anomaly equation leads to $\delta\rho=\frac{G}{2\pi}\delta a_z$, with $G=\frac{eB_0}{h}$ the Landau level degeneracy. Therefore the change of $a_z$ leads to a charge density modulation proportional to it. To understand this equation we consider
the Landau level spectrum for the four band model (\ref{eq:MR_Ham1}), as is shown in Fig \ref{fig:setup} (b). It is important to note that the two zeroth Landau levels with opposite chirality have the same spin polarization. Consequently, the exchange coupling of $\vec{a}=a_z\hat{z}$ with the zeroth Landau level states is equivalent to a scalar potential, which shifts the chemical potential and leads to the change of charge density.

Since this term couples charge density and magnetization, it leads to an interesting physical consequence of the hybridization between the plasmon and magnon modes.
The effective action of the present system can be given by
$S_{A}+S_{a}+S_{RPA}$, where $S_{A}=\int dkd\omega A_0(q)G_{A0}^{-1} A_0(-q)$, $S_a=\int dkd\omega a_z(q)G_a^{-1}a_z(-q)$ and
$S_{RPA}=\int dkd\omega \left[ (A_0(q)+a_z(q))\Pi(A_0(-q)+a_z(-q)) \right]$ with $ q=(\omega,k)$. Here $a_z$ has been rescaled to have the same
dimension as $A_0$. 
$S_A$ describes the dynamics of the scalar potential $A_0$ with $G_A^{-1}\sim v_c^2k^2$ with photon velocity $v_c$, $S_a$ describes
the dynamics of $a_z$ field with $G_a^{-1}\sim \omega^2-v_s^2k^2-\omega_0^2$ where $\omega_0$ gives the excitation gap and $v_s$ is the magnon velocity,
and $S_{RPA}$ gives the effective action after integrating out the interacting fermions with the random phase approximation, where
$\Pi\sim -\frac{\omega_p^2}{\omega^2}v_c^2k^2$\cite{nagaosa1999} ($\omega_p$ is the plasmon frequency). Such a hybridization is shown schematically in Fig \ref{fig:setup} (a). To see the effect on the dynamics of the magnon described by $a_z$, we integrate out $A_0$ field and obtain $S_{eff}=\int dkd\omega a_z(q)\left( G^{-1}_a+\frac{\Pi G_A^{-1}}{\Pi+G_A^{-1}} \right)a_z(-q)$.  The corresponding correlation function is given by $\langle a_z(q)a_z(-q)\rangle\sim \left( \omega^2-v_s^2k^2-\omega_0^2-\frac{\omega_p^2v_c^2k^2}{\omega^2-\omega_p^2+i\eta}+i\eta \right)^{-1}$, which
corresponds to the spin susceptibility. As plotted in Fig \ref{fig:setup} (c), the correlation function has two poles, of which one corresponds to the intrinsic magnon excitation with the frequency around $\omega_0$,
while the other one only appears for finite $k$ with the intensity proportional to $k^2$ and is induced by the plasmons with frequency around $\omega_p$.
The plasmon frequency can be estimated as $\sim35meV$ for Weyl fermions \cite{dassarma2009} with dielectric constant $\sim100$, Fermi velocity $\sim6.85\times 10^5 m/s$,
and electron density $\sim10^{19}$ cm$^{-3}$. Such an additional mode in magnon spectrum can be observed in neutron scattering experiments and compared with the plasmon frequency $\omega_p$ determined by reflection spectroscopy. 

The authors would like to thank J. Jain, K. Sun, C.K. Xu, Y.S. Wu and S.C. Zhang for helpful discussions. This work is
supported by Tsinghua Education Foundation North America (P.Y.), and the Defense Advanced Research Projects Agency Microsystems Technology Office,
MesoDynamic Architecture Program (MESO) through the contract number N66001-11-1-4105 (X.L.Q). 
This work was supported in part by the National Science Foundation under Grant No. NSF PHY11-25915 when the authors participated in the
KITP program Topological Insulators and Superconductors. We thank KITP for hospitality.

\appendix 
\section{Numerical method for the calculation of the chiral mode in the ferromagnetic vortex core}
In the appendix, we describe our numerical method for the calculation of energy dispersion and eigen wavefunction for the ferromagnetic vortex configuration.
We start from the four band model (4) in the main text and in the cylinder coordinate $(r,\theta,z)$, the Hamiltonian takes the form of
\begin{eqnarray}
	&&H=H_0+H_1\label{eq:AP_Ham1}
\end{eqnarray}
\begin{eqnarray}
	&&H_0=\mathcal{M}(\vec{k})\tau_z+B_0k_z\tau_y+A_0(k_y\sigma_x\tau_x-k_x\sigma_y\tau_x)\nonumber
\end{eqnarray}
\begin{eqnarray}
	&&=\left(
	\begin{array}{cccc}
		\mathcal{M}(\vec{k})&0&-iB_0k_z&iA_0k_-\\
		0&\mathcal{M}(\vec{k})&-iA_0k_+&-iB_0k_z\\
		iB_0k_z&iA_0k_-&-\mathcal{M}(\vec{k})&0\\
		-iA_0k_+&iB_0k_z&0&-\mathcal{M}(\vec{k})
	\end{array}
	\right)\nonumber
\end{eqnarray}
\begin{eqnarray}
	&&H_1=-W_0\sin\theta\sigma_x\tau_z+W_0\cos\theta\sigma_y\tau_z+U_0\sigma_z\nonumber\\
	&&=\left(
	\begin{array}{cccc}
		U_0&-iW_0e^{-i\theta}&0&0\\
		iW_0e^{i\theta}&-U_0&0&0\\
		0&0&U_0&iW_0e^{-i\theta}\\
		0&0&-iW_0e^{i\theta}&-U_0
	\end{array}
	\right)\nonumber
\label{}
\end{eqnarray}
where we have $\partial_x=\cos\theta\partial_r-\frac{\sin\theta}{r}\partial_\theta$ and
$\partial_y=\sin\theta\partial_r+\frac{\cos\theta}{r}\partial_\theta$, therefore
$k_-=k_x-ik_y=-i\partial_x-\partial_y=-i\left( \cos\theta\partial_r-\frac{\sin\theta}{r}\partial_\theta \right)
-\left( \sin\theta\partial_r+\frac{\cos\theta}{r}\partial_\theta \right)=-ie^{-i\theta}\partial_r
-\frac{e^{-i\theta}}{r}\partial_\theta$, $k_+=-i\partial_x+\partial_y=-i\left( \cos\theta\partial_r-\frac{\sin\theta}{r}\partial_\theta \right)
+\left( \sin\theta\partial_r+\frac{\cos\theta}{r}\partial_\theta \right)=-ie^{i\theta}\partial_r
+\frac{e^{i\theta}}{r}\partial_\theta$ and
$k_x^2+k_y^2=-\left( \frac{\partial^2}{\partial r^2}+\frac{1}{r}\frac{\partial}{\partial r}+\frac{1}{r^2}\frac{\partial}{\partial \theta^2}\right)$.
The above Hamiltonian has in-plane rotation symmetry along z axis and the corresponding total angular momentum can be defined as $J_z=L_z+\frac{1}{2}\sigma_z$
where $L_z=-i\frac{\partial}{\partial\theta}$ and the Pauli matrix $\sigma_z$ denotes the spin part. With the in-plane rotation symmetry, the wavefunction ansatz can be taken as
$\tilde{\psi}(r,\theta)=\left[ e^{in\theta}f_1(r),e^{i(n+1)\theta}f_2(r),e^{in\theta}f_3(r),e^{i(n+1)\theta}f_4(r) \right]^T$
where the total angular momentum $J_z=n+\frac{1}{2}$. The Hamiltonian is changed to
\begin{widetext}\begin{eqnarray}
	\tilde{H}=\left(
	\begin{array}{cccc}
		\tilde{\mathcal{M}}(n)+U_0&-iW_0&-iB_0k_z&A_0\left( \partial_r+\frac{n+1}{r} \right)\\
		iW_0&\tilde{\mathcal{M}}(n+1)-U_0&A_0\left( -\partial_r+\frac{n}{r} \right)&-iB_0k_z\\
		iB_0k_z&A_0\left( \partial_r+\frac{n+1}{r} \right)&-\tilde{\mathcal{M}}(n)+U_0&iW_0\\
		A_0(-\partial_r+\frac{n}{r})&iB_0k_z&-iW_0&-\tilde{\mathcal{M}}(n+1)-U_0
	\end{array}
	\right)
	\label{eq:H1}
\end{eqnarray}\end{widetext}
where $\tilde{\mathcal{M}}(n)=M_0+M_1k_z^2-M_2\left( \frac{\partial^2}{\partial r^2}+\frac{1}{r}\frac{\partial}{\partial r}-\frac{n^2}{r^2} \right)$
and the wave function is now given by $\tilde{\psi}=\left[ f_1(r),f_2(r),f_3(r),f_4(r) \right]^T$. Let's introduce the new wave function $\psi$
as $\tilde{\psi}=\frac{1}{\sqrt{r}}\psi$, then the normalization relation $\int rdrd\theta |\tilde{\psi}|^2=1$ is changed to $\int drd\theta |\psi|^2=1$,
and the effective Hamiltonian is rewritten as
\begin{widetext}\begin{eqnarray}
	&&H=\left(
	\begin{array}{cccc}
		\mathcal{M}(n)+U_0&-iW_0&-iB_0k_z&A_0\left( \partial_r+\frac{n+1/2}{r} \right)\\
		iW_0&\mathcal{M}(n+1)-U_0&A_0\left( -\partial_r+\frac{n+1/2}{r} \right)&-iB_0k_z\\
		iB_0k_z&A_0\left( \partial_r+\frac{n+1/2}{r} \right)&-\mathcal{M}(n)+U_0&iW_0\\
		A_0(-\partial_r+\frac{n+1/2}{r})&iB_0k_z&-iW_0&-\mathcal{M}(n+1)-U_0
	\end{array}
	\right)
	\label{eq:H2}
\end{eqnarray}\end{widetext}
where $\mathcal{M}(n)=M_0+M_1k_z^2-M_2\left( \frac{\partial^2}{\partial r^2}-\frac{n^2-1/4}{r^2} \right)$.
This Hamiltonian can be written in a compact form
\begin{widetext}\begin{eqnarray}
	&&H=\left[ M_0+M_1k_z^2-M_2\left( \frac{\partial^2}{\partial r^2}-\frac{(n+1/2)^2}{r^2} \right) \right]\tau_z-M_2\frac{n+1/2}{r^2}\sigma_z\tau_z\nonumber\\
	&&+B_0k_z\tau_y+iA_0\frac{\partial}{\partial r}\sigma_y\tau_x+A_0\frac{n+1/2}{r}\sigma_x\tau_x+W_0\sigma_y\tau_z+U_0\sigma_z.
	\label{eq:H3}
\end{eqnarray}\end{widetext}

We can discretize the Hamiltonian (\ref{eq:H2}) and solve the eigenstate problem for the radial equation numerically.
The corresponding result is shown in Fig 2 of the main text.
For $n=0$, we indeed find two gapless modes with the opposite velocities along z direction, and these two gapless modes are spatially separated
with one wave function mainly staying at $r=0$ and the other one at $r=R$, as shown by the red and black lines in Fig 2 (a) and (b) of the main text. However with increasing
n, a gap is opened between the two low energy modes, as shown in Fig 2 (c) in the main text.
To get more analytical understanding of the radial equation, we consider the $r\rightarrow \infty$ limit with
$U_0=0$, $k_z=0$, where the radial Hamiltonian is simplified as $H=\left( M_0-M_2\frac{\partial^2}{\partial r^2}\right)\tau_z
+W_0\sigma_y\tau_z+iA_0\sigma_y\tau_x\frac{\partial}{\partial r}$. With the wave function ansatz $\psi\sim e^{\lambda r}\phi$, we obtain the equation
$A_0\lambda\phi=\left[ \left( M_0-M_2\lambda^2 \right)\sigma_y\tau_y+W_0\tau_y \right]\phi$ for the zero modes. Since $\left[ \sigma_y\tau_y,\tau_y \right]=0$,
we can take the common eigen-states of $\sigma_y\tau_y$ and $\tau_y$ for $\phi$, $\sigma_y\tau_y\phi_{ts}=t\phi_{ts}$ and $\tau_y\phi_{ts}=s\phi_{ts}$,
then the wave function can be expressed as $\psi=\sum_{\alpha,t,s}c_{\alpha,ts}e^{\lambda_{\alpha}(t,s)r}\phi_{t,s}$, with $\lambda$ given by
$\lambda_{\alpha}(t,s)=\frac{-tA_0+\alpha\sqrt{A_0^2+4M_2\left( tsW_0+M_0 \right)}}{2M_2}$. The existence of the edge mode requires $\lambda_+(+,+)
\lambda_-(+,+)>0$ or $\lambda_+(+,-)\lambda_-(-,-)>0$, leading to the following different regimes: in the normal regime $M_0M_2>0$,
the system has no zero mode when $|W_0|<|M_0|$ and one zero mode when $|W_0|>|M_0|$, while in the inverted regime $M_0M_2<0$,
the system has one zero mode when $|W_0|>|M_0|$ and two zero modes when $|W_0|<|M_0|$. Taking into account the $k_z$ dependent term,
it turns out that one zero mode case corresponds to the 1D chiral state and two zero modes case is
the 1D helical state. However since time reversal is broken in the present system, the helical state is not protected and can be gapped.
Therefore the only robust state is the chiral state when $|W_0|>|M_0|$.
We emphasize that the transition at $|W_0|=|M_0|$ exactly corresponds to the condition for the appearance of the gapless Weyl fermions for the uniform magnetization.
For the finite r, the terms proportional to $\frac{1}{r}$ and $\frac{1}{r^2}$ will push the chiral mode around r=0 outwards,
thus with increasing the angular momentum number n, the wave function of the chiral mode near $r=0$ extends to the large r region
and mixes with the chiral mode at $r=R$, opening a gap.


\end{document}